\documentclass[rnote]{aa}

\usepackage{natbib}
\usepackage{longtable}
\usepackage{lscape}
\usepackage{amsmath}
\usepackage{txfonts}
\usepackage{color}
\usepackage{url}
\usepackage{xspace}

\usepackage{graphicx}
\bibpunct{(}{)}{;}{a}{}{,}

\newcommand{\kev}{keV\xspace}
\newcommand{\ergs}{erg$\,$s$^{-1}$\xspace}
\newcommand{\ergscm}{erg$\,$s$^{-1}$cm$^{-2}$\xspace}
\newcommand{\xmm}{\textsl{XMM-Newton}\xspace}
\newcommand{\rosat}{\textsl{ROSAT}\xspace}
\newcommand{\asca}{\textsl{ASCA}\xspace}

\newcommand{\chandra}{\textsl{Chandra}\xspace}

\begin{document}
   \title{XMMU~J134736.6+173403: an eclipsing LMXB  in quiescence
   or a peculiar AGN?}

   \author{S. Carpano
          \inst{1}
          \and
	  B. Altieri
          \inst{2}
          \and
	  A.~R. King
          \inst{3}
          \and
	  A. Nucita
          \inst{1}
          \and
	  P. Leisy
          \inst{4}}

   \offprints{S. Carpano, e-mail: scarpano@sciops.esa.int}
   \institute{XMM-Newton Science Operations Centre, ESAC, ESA, PO Box 50727, 28080 Madrid, Spain
              \and
	      Herschel Science Centre, ESAC, ESA, PO Box 50727, 28080 Madrid, Spain
              \and
	     Department of Physics and Astronomy, University of Leicester, Leicester LE1 7 RH, UK
              \and
	     Isaac Newton Group of Telescopes, Apartado de Correos 321, E-38700 Sta. Cruz de La Palma, Spain; Instituto de Astrofísica de Canarias, E-38205 La Laguna, Tenerife, Spain}
 
   \date{Submitted: 8 November 2007; Accepted: 8 January 2008}
   
   \abstract{}{We report the discovery of a peculiar object observed serendipitously with \xmm. We present its timing and spectral properties 
   and investigate its optical counterpart. }{The light curve of the X-ray source, its spectrum,
   and the spectrum of the best optical counterpart are presented and analyzed.}{The X-ray flux  decreases by a factor of 6.5 within 1\,h and
   stays in a low state 
   for at least 10\,h, thereby suggesting the presence of an eclipse. The spectrum is very soft, a power law with a slope of $\Gamma\sim2.8$, and does not change
   significantly before and after the flux drop. The source is spatially coincident within few arc-seconds with a Seyfert~2 galaxy belonging to a galaxy
   pair.}{Although the background AGN seems the best counterpart, neither the temporal nor the spectral properties of the X-ray source are
   compatible with it. We investigate the possibility of having a foreground low-mass X-ray binary in quiescence, where the companion is not 
   detected in the optical wavelength.}


\maketitle
%

\section{Introduction}
In this paper we report the discovery of an X-ray source for which the flux drops abruptly by a factor of 6.5 within one
hour, and stays in this low state for at least 10 hours.
The nature of the source described in this paper is very ambiguous. The X-ray light curve suggests the presence of an 
eclipse and hence a binary system. The detected optical counterpart is a pair of galaxies, one of which is a Seyfert~2 AGN.
The X-ray spectrum is also very soft so as to be compatible with such an AGN. We suggest the possibility that we have a foreground low-mass
X-ray binary in quiescence for which the optical counterpart is below detectability, but we do not exclude the  presence of 
a peculiar AGN. In this section we introduce the general X-ray properties both of low-mass X-ray binaries in quiescence and of Seyfert~2 galaxies.

Low-mass X-ray binaries (LMXBs) are composed of a compact object, either a neutron star (NS) or a black hole (BH), accreting matter from a
companion star with masses $\lesssim1$\,M$_{\odot}$. These objects have typical X-ray luminosities of about
10$^{37}$--10$^{38}$\,\ergs but can drop down to 10$^{31}$--10$^{33}$\,\ergs during quiescence. 
At high luminosities, the disk is optically thick and geometrically thin, while a hot optically thin, advection-dominated accretion disk is
expected for lower luminosities.
In the advection-dominated accretion flow (ADAF) model, the radiative efficiency is very low ($\sim$10$^{-4}$--10$^{-3}$) and most of the
gravitational energy is stored as internal energy and advected towards  the compact object with little X-ray emission \citep{Narayan1996}.
A hot quasi-spherical flow is expected in the inner regions producing X-rays \citep{Narayan1996}.

The spectrum of NS-LMXBs in quiescence displays a soft (kT$\sim$0.1\,\kev) thermal component \citep{McClintock2004a},
probably associated with the stellar surface. Indeed, the observed X-ray luminosities, temperatures, and distance of these systems indicate that
the thermal emission comes from a source with a radius of $\sim$10\,km.
On the other hand, for BH-LMXBs in quiescence, no thermal emission has been reported for the 15 sources reported by  
\cite{McClintock2004b}, since they possess an event horizon. Their spectra are well-fitted by a single power law with a photon index
1.5$<\Gamma<$2.1 \citep{McClintock2004b}.

The brightest and best-studied NS-LMXBs in quiescence are Cen~X-4 and Aql~X-1. Spectral and temporal analysis of these sources during quiescence are
reported, for example, by \cite{Campana2004} and \cite{Campana2003}, respectively. Both sources display a thermal component and a power law, as well as
short-timescale temporal variability \citep{Campana2003,Rutledge2002}. The X-ray Nova V~404 Cygni is, on the other hand, the brightest
stellar-mass BH 
in quiescence. A recent review of its properties is reported by \cite{Bradley2007}: the spectrum in quiescence is typically a power law with a $\Gamma\sim2$, 
consistent with with what is expected for the advection-dominated accretion flow. In the low state, the source is also variable on a short time scale 
up to a factor of $\gtrsim$20 during a 60 ks observation \citep{Hynes2004}.

We review in this paragraph some of the X-ray properties of Seyfert~2 galaxies. In the unified model for AGN \citep{Antonucci1993}, Seyfert~2 galaxies 
are identical to the Seyfert~1, but observed at high inclination
angle. The nucleus of Seyfert~2 galaxies is therefore not directly visible,  and shows much higher column densities than in
Seyfert~1 objects, blocking the soft X-ray flux below 2\,\kev. \cite{Risaliti1999} show that all sources of their sample of Seyfert~2 galaxies have 
$N_\text{H}\ge10^{22}\,\text{cm}^{-2}$, and 75\% $N_\text{H}\ge10^{23}\,\text{cm}^{-2}$. Their spectra are relatively hard. \cite{Moran2001} show a composite spectrum
for 29 Seyfert~2 galaxies observed with \asca. The continuum is described by a double power law component: an unabsorbed component (energy index $\alpha$=0.84) dominates
below 3\,\kev, and at higher energies the spectrum is dominated by a heavily absorbed component ($N_\text{H}=3\times10^{23}\,\text{cm}^{-2}$,
$\alpha$=0.46). An Fe
K$\alpha$ line is present at 6.4\,\kev, with an equivalent width of 420\,eV. Small-amplitude, short-timescale variability is observed
for Seyfert~2 galaxies, as \cite{Awaki2006}. These authors report variation of a factor less than two for
13 Seyfert~2 galaxies observed with \xmm, on a time scale of a few thousand seconds.

The paper is organized at follows. Section~\ref{sec:obs} 
describes the \xmm and optical observation and data reduction. In Sect.~\ref{sec:time}, we present
timing and spectral analysis of the source, reserving discussion for Sect.~\ref{sec:conc}.  


\section{Observation and data reduction}
\label{sec:obs}
\subsection{\xmm data}
XMMU~J134736.6+173403 was observed serendipitously by \xmm on 2003 June 24 for 64\,ksec, the target being 
Tau~Boo. The  EPIC-MOS \citep{Turner} were operated in large window mode and EPIC-pn \citep{Strueder} cameras  in
full frame mode with the thick filter for all instruments. The optical monitor was blocked throughout the full
observation. After screening the MOS data for proton flares using standard 
procedures\footnote{\url{http://xmm.esac.esa.int/external/xmm_user_support/documentation/sas_usg/USG/}},
a total of 60 and 56\,ksec of low-background emission remained for the MOS and pn, respectively. 

Using the \xmm Software Analysis System (SAS) \texttt{edetect\_chain} task, which performs maximum-likelihood 
source detection, XMMU~J134736.6+173403, observed at $\sim8'$ off-axis,
was detected with a maximum likelihood of $3\times10^4$. The best-fit coordinates were
$\alpha_\text{J2000}=13^\text{h}47^\text{m} 36\fs{}6$ and
$\delta_\text{J2000}=+17^\circ 34' 02\farcs 8$ with a statistical error of $0\farcs 1$. 
Since the source is off-axis, however, the uncertainty on the statistical error might be much greater than
what is provided by \texttt{edetect\_chain}. Combining this error to the systematical shift expected between 
X-ray data and optical counterpart (typically few arcsec), the total estimated error is  about $3\farcs$

\subsection{Optical counterpart spectra}
Low-resolution spectroscopy was performed with ISIS
at the William Herschel Telescope (ING, La Palma) in the night  2007 June 14 in service mode.
The slit was aligned east-west to get the spectra of the galaxy and the AGN in the same frames.
The blue and red spectra were acquired simultaneously on two different CCDs,
thanks to a dichroic system.
Two short exposures of 10\,mn each were made to avoid cosmic rays. 

The data reduction was performed in a standard way  for both the blue and red spectrum parts.
We wrote a dedicated and semi-automatic MIDAS batches for ISIS long-slit spectra essentially using
the {\it LONG} package.
We first performed the standard reductions, i.e. a bias substraction,
a flat-fielding, and a cosmic rejection.
Then we  extracted the arcs at the object position, summed them if more than one, and
from them we identified some lines and automatically performed a wavelength calibration.
All the objects were calibrated in wavelength, corrected from the mean
extinction, and calibrated in fluxes with 3 standards stars taken during the night.
This instrumental response is very stable in time, regular checks are done monthly.
Finally, from these 2D images we  extracted the objects with an optimized
spectrum extraction batch and a line-fitting to measure the line intensities and the
redshift.


\section{Timing and spectral analysis of XMMU~J134736.6+173403}
\label{sec:time}

\begin{figure}
  \resizebox{\hsize}{!}{\includegraphics{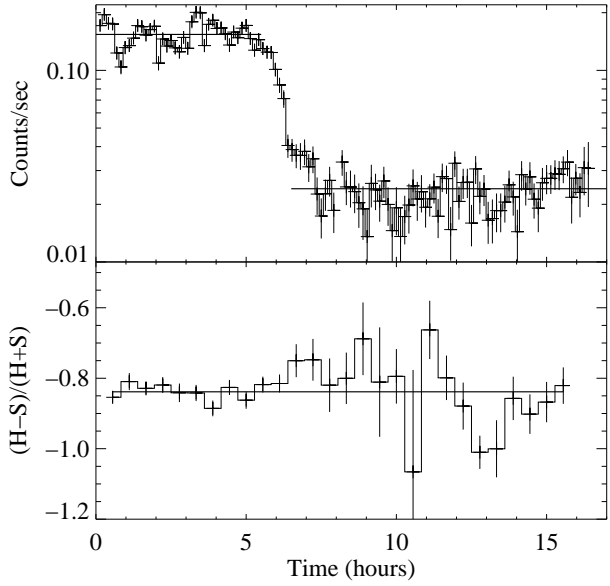}}
 \caption{Mean \xmm EPIC-MOS and pn background-subtracted light curve of XMMU~J134736.6+173403 (top),
 and the corresponding hardness-ratio light curve (bottom).  
 The straight lines show the average of the respective portion of the curves.}
 \label{fig:light}
\end{figure}

The combined EPIC-MOS and pn background-subtracted light curve is shown in Fig.~\ref{fig:light} (top), with a time bin
size of 500\,sec. The corresponding hardness-ratio light curve is given at the bottom, with a time bin
size of 2000\,sec. Hard and soft bands are given by H=2--10\,\kev and S=0.2--2\,\kev, respectively.
Times are given in hours from the start of the observation.
Periods of high background at the end of the observation were excluded from the data.
The straight lines show the average of the respective portion of the curves.
In the first curve the flux drops by a factor of 6.5 within 1\,hr. 
We searched for
periodicities between 10\,s and 6\,hr in the light curve before the dip where the flux is visibly variable. 
We found a significant modulation at 1.8\,h, but more data are needed to confirm the periodicity.
On the other hand, no periodic signal was found in the low state.
The fractional root mean-square variability amplitude, defined for example in \cite{Vaughan2003}, $F_\text{var}$, is 
13.7$\pm$1.2\% before the flux drop and 13.2$\pm$3.8\% after it, showing that the short-term variability is roughly constant during the observation.
The hardness-ratio light curve does not vary after the flux decline either.

\begin{figure}
  \resizebox{\hsize}{!}{\includegraphics[bb=113 44 563 708,clip=true,angle=-90]{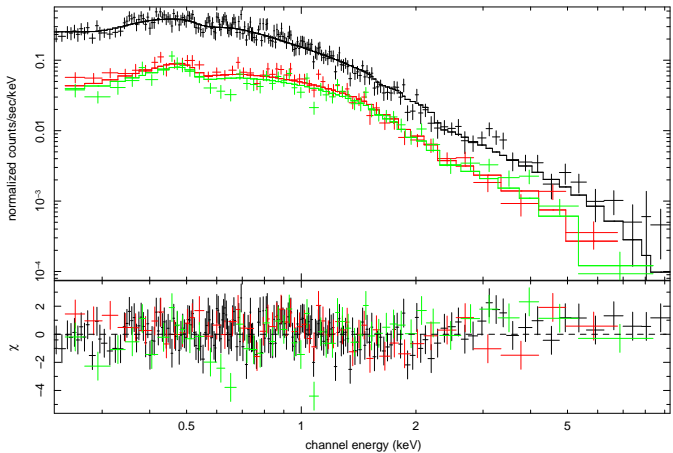}}
 \caption{\xmm EPIC-pn and MOS spectra of  XMMU~J134736.6+173403 in the high state, fitted with an absorbed power law. 
 Bottom: residuals expressed in $\sigma$.}
 \label{fig:spec}
\end{figure}

Figure~\ref{fig:spec} shows the pn and MOS spectra of the source. The data are
binned to have at least 25 counts in each energy bin. The source is described well by a power law model, yielding
$\chi^2_{\nu}/\nu=1.18$ in the high state and $\chi^2_{\nu}/\nu=0.85$ in the low. Adding another component does not 
improve the fit.
The best-fitting parameters of the absorbed power law  model are shown in Table~\ref{tab:spec_fit}. Here, $N_\text{H}$ is the
equivalent column density of neutral hydrogen and $\Gamma$ the photon
index. The corresponding 0.2--10\,\kev flux is shown in the last 
row. Uncertainties are given at a 90\% confidence level.

\begin{table}
 \centering
 \caption{Results of the spectral fits for XMMU~J134736.6+173403, using an
   absorbed power law model
   (\texttt{phabs*power}, in XSPEC). } 
  \label{tab:spec_fit}
 \begin{tabular}{lll}
 \hline  
  & High State & Low state\\
 \hline
 $N_\text{H}(\times 10^{20}\,\text{cm}^{-2})$& 4.73$^{+0.90}_{-0.83}$& 3.13$^{+2.12}_{-1.75}$ \\[3pt]
 $\Gamma$& $\,2.81^{+0.07}_{-0.07}\,$ & $\,2.72^{+0.19}_{-0.15}\,$  \\[3pt]
 $F_{0.2-10\,\text{\kev}}\times 10^{-12}$ (\,cgs) & 1.33$^{+0.05}_{-0.05}$ & 0.24$^{+0.03}_{-0.02}$ \\[3pt]
 \hline
   \end{tabular}
\end{table}

We searched for X-ray data for the source from other X-ray satellites: no observation has been performed with \chandra, and the source is
 detected in neither the
all-sky survey nor in a short (600\,s) PSPC observation from the \rosat satellite, where the upper limit flux was of $3\times10^{-13}$ and
$8\times10^{-13}$\,\ergscm, respectively. The source was, however, visible in a $\sim$80\,ks \asca
observation and reported in the catalogue of \asca sources \citep{Ueda2005}. Their flux and hardness-ratios  are compatible
to the mean value measured by \xmm. We extracted the \asca light curve, and there does not seem to be any flux drop like the one observed in the
\xmm data, although the count rate is very low (5\,cts ks$^{-1}$).


\section{The optical counterparts}
\label{sec:optic}

\begin{figure}
  \resizebox{\hsize}{!}{\includegraphics[bb=15 69 764 633,clip=true]{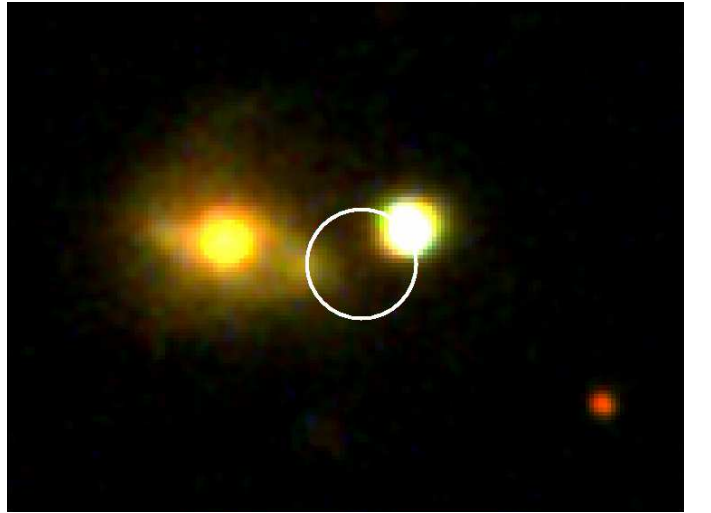}}
 \caption{Color-composite image of the optical counterpart from the Sloan Digital Sky Survey (SDSS). The red color is associated with
 the R band, the green color  with the G band, and the blue color  with the U band.
 The circle is centered on the X-ray source and the radius is  3$''$.}
 \label{fig:image}
\end{figure}

\begin{figure}
  \resizebox{\hsize}{!}{\includegraphics{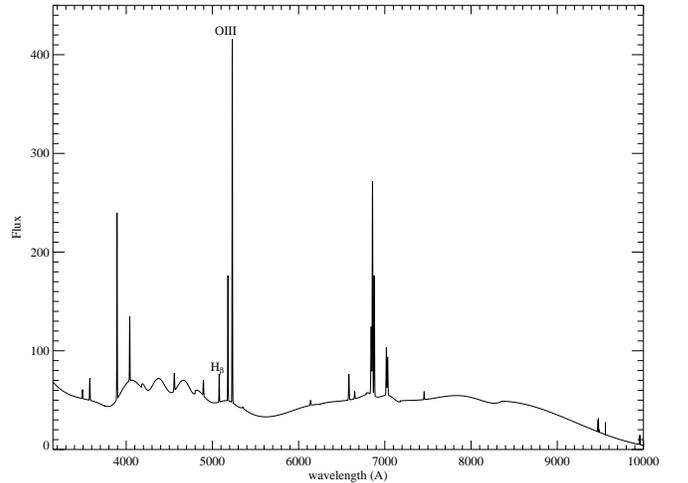}}
 \caption{Optical spectrum of the best optical counterpart of  XMMU~J134736.6+173403. The width of the H$_\beta$ line, and the ratio between
 O$_\text{III}$ over H$_\beta$  ($<$3) is consistent with the spectrum of a Seyfert~2 galaxy.}
 \label{fig:optic_spec}
\end{figure}

Figure~\ref{fig:image} shows the optical counterpart image of the X-ray source. The circle is centered on the position of the X-ray source and the radius 
is 3$''$.  Both sources are extra-galactic objects with a redshift of z=0.045. From the image and the spectrum, the source on
the left is a normal galaxy with some emission lines (not shown in this paper). The object on the right has a spectrum typical of a
Seyfert~2 galaxy (narrow H$_\beta$ line and O$_\text{III}$/H$_\beta<$3, see Fig.~\ref{fig:optic_spec}). At this distance the sources are separated by 10\,kpc 
(corresponding to 10$''$), suggesting that they form a pair of galaxies, maybe in interaction.
The Seyfert~2 galaxy is also associated with a radio source with an intensity of 17.91 mJy at 1.4\,Ghz \citep{Becker2003}.

\section{Discussion}
\label{sec:conc}

We reported the discovery of an X-ray source that presents a  flux drop of a factor of 6.5 within 1\,h and stays in a low state for at least 10\,h,
suggesting  an eclipse. From optical archived images, there are two counterparts: the first  an extended normal galaxy and
the second a point source. A radio source is also associated to this last one. We got the optical spectrum of the two objects from the ING Telescope in La
Palma. The extended object is indeed a redshifted object with emission lines consistent with a normal or slightly active galaxy. Surprisingly, the second
object appeared at the same redshift (z=0.045) and has a spectrum typical for a Seyfert~2 galaxy. At that distance the objects are separated by 10\,kpc,
suggesting that they form a pair of galaxies, maybe in interaction.

Within this context, the X-ray source would be associated with the Seyfert~2 galaxy (generally strong X-ray sources) rather than the normal 
galaxy. At this distance, the observed luminosity in the 0.2-10\,\kev band is 6.5$\times10^{42}$\,\ergs in the high state and 1$\times10^{42}$\,\ergs in the
low state. This kind of luminosity is typical of Seyfert~2 galaxies. The temporal and spectral properties of the X-ray object are
unexpected, however, for such an AGN. A flux drop of a factor of 6.5 within 1\,h has never been reported so far, especially for Seyfert~2 galaxies where the galaxy
is observed edge-on. The spectrum, furthermore, is extremely soft for such a galaxy and does not present any K$\alpha$ line at 6.4\,\kev.

We thus believe that the source is a foreground object within the line-of-sight of this pair of galaxies. The detection limit of the Sloan Digital Sky
survey is  23.3 mag and 23.1 mag in the green and red bands, respectively \citep{Willman2002}. Since the size of the Galaxy towards the source is 
 about 5\,kpc, the companion must be type M2, M$_\text{V}$=9.9 \citep{Cox2000} or later to explain the non-detectability.
Assuming the object is at 5\,kpc, the luminosity is  4$\times$10$^{33}$\,\ergs and becomes 6$\times$10$^{32}$\,\ergs if the source is at only 2\,kpc
(implying therefore a M4/M5 dwarf companion). This is well within the range of what is expected for low-mass X-ray binaries in quiescence and is consistent
with  these objects having  K or M companion stars. We exclude the presence of a cataclysmic variable, since their orbital periods are 
only a few hours, while here, if the flux drop is associated with an eclipse, it must be at least $\sim$1 day.
We note that the spectrum is a bit too soft ($\Gamma\sim2.8$) for an BH-LMXB in quiescence and that a power law continuum in unexpected
for a thermal emission from a neutron star surface.

If the X-ray source is a foreground object, then the X-ray flux of the background AGN should also be detected and both are probably confused.
 A deep \chandra observation of the system might represent the only solution for disentangling  the LMXB picture or the peculiar AGN scenario.
Thanks to its better spatial resolution, we might be able to resolve 2 hypothetical point sources in the first scenario, while one single source,
associated with the AGN, should be observed in the second case.

\begin{acknowledgements}
  This paper is based on observations obtained with \textsl{XMM-Newton}, an ESA
  science mission with instruments and contributions directly funded
  by ESA Member States and NASA, and on observations made with the WHT telescope operated    
  on the island of La Palma by the Isaac Newton Group in the Spanish
  Observatorio del Roque de los Muchachos of the Instituto de
  Astrofisica de Canarias.
  We acknowledge the following persons (in alphabetic order) for their help in the interpretation 
  of the source nature: L.~Ballo, A.-L.~Longinotti, M.~Guainazzi, A.~Pollock, A.~Read, R.~Saxton, M.D.~Trigo.

  Funding for the SDSS and SDSS-II has been provided by the Alfred P. Sloan Foundation, the Participating 
  Institutions, the National Science Foundation, the U.S. Department of Energy, the National Aeronautics 
  and Space Administration, the Japanese Monbukagakusho, the Max Planck Society, and the Higher Education 
  Funding Council for England. The SDSS Web Site is \url{http://www.sdss.org/}.

    The SDSS is managed by the Astrophysical Research Consortium for the Participating Institutions. 
    The Participating Institutions are the American Museum of Natural History, Astrophysical Institute 
    Potsdam, University of Basel, University of Cambridge, Case Western Reserve University, University of 
    Chicago, Drexel University, Fermilab, the Institute for Advanced Study, the Japan Participation Group, 
    Johns Hopkins University, the Joint Institute for Nuclear Astrophysics, the Kavli Institute for Particle 
    Astrophysics and Cosmology, the Korean Scientist Group, the Chinese Academy of Sciences (LAMOST), Los Alamos 
    National Laboratory, the Max-Planck-Institute for Astronomy (MPIA), the Max-Planck-Institute for Astrophysics 
    (MPA), New Mexico State University, Ohio State University, University of Pittsburgh, University of Portsmouth, 
    Princeton University, the United States Naval Observatory, and the University of Washington.

\end{acknowledgements}


\begin{thebibliography}{30}
\expandafter\ifx\csname natexlab\endcsname\relax\def\natexlab#1{#1}\fi

\bibitem[{Antonucci}(1993)]{Antonucci1993}
{Antonucci}, R. 1993, \araa, 31, 473

\bibitem[{Awaki}  {et~al.}(2006)]{Awaki2006}
{Awaki}, H., {Murakami}, H., {Ogawa}, Y. \& {Leighly}, K.~M. 2006, \apj, 645, 928

\bibitem[{Becker}  {et~al.}(2003)]{Becker2003}
{Becker}, R.~H.,  {Helfand}, D.~J., {White}, R.~L., {Gregg}, M.~D. \& {Laurent-Muehleisen}, S.~A. 2003, 
VizieR Online Data Catalog, 8071

\bibitem[{Bradley}  {et~al.}(2007)]{Bradley2007}
{Bradley}, C.~K., {Hynes}, R.~I., {Kong}, A.~K.~H., {Haswell}, C.~A. et al. 2007, \apj, 667, 427

\bibitem[{{Campana} \& {Stella}(2003)}]{Campana2003}
{Campana}, S. \& {Stella}, L. 2003, \apj, 597, 474

\bibitem[{Campana}  {et~al.}(2004)]{Campana2004}
{Campana}, S.,  {Israel}, G.~L.,  {Stella}, L.,  {Gastaldello}, F. \& {Mereghetti}, S. 2004, \apj, 601, 474

\bibitem[{Cox}(2000)]{Cox2000}
{Cox}, A.~N. 2000, Allen's astrophysical quantities, 4th ed.~Publisher: New York: AIP Press; Springer

\bibitem[{Fabbiano}  {et~al.}(2003)]{Fabbiano2003}
{Fabbiano}, G., {King}, A.~R., {Zezas}, A., {Ponman}, T.~J., {Rots}, A. \&
{Schweizer}, F. 2003, \apj, 591, 843

\bibitem[{Hynes}  {et~al.}(2004)]{Hynes2004}
{Hynes}, R.~I., {Charles}, P.~A., {Garcia}, M.~R., {Robinson}, E.~L. et al. 2004, \apjl, 611, L125

\bibitem[{McClintock}  {et~al.}(2004)]{McClintock2004a}
{McClintock}, J.~E., {Narayan}, R. \& {Rybicki}, G.~B. 2004, \apj, 615, 402

\bibitem[{{McClintock} \& {Remillard}(2004)}]{McClintock2004b}
{McClintock}, J.~E. \& {Remillard}, R.~A. 2004, in Compact Stellar X-Ray Sources, 
ed. W.~H.~G. Lewin \& M. van der Klis (cambridge: Cambridge Univ. Press), astro-ph/0306213

\bibitem[{Moran}  {et~al.}(2001)]{Moran2001}
{Moran}, E.~C., {Kay}, L.~E., {Davis}, M., {Filippenko}, A.~V.\& {Barth}, A.~J. 2001, \apjl, 556, L75

\bibitem[{Narayan}  {et~al.}(1996)]{Narayan1996}
{Narayan}, R., {McClintock}, J.~E. \& {Yi}, I. 1996, \apj, 457, 821

\bibitem[{Risaliti}  {et~al.}(1999)]{Risaliti1999}
{Risaliti}, G., {Maiolino}, R. \& {Salvati}, M. 1999, \apj, 522, 157

\bibitem[{Rutledge}  {et~al.}(2002)]{Rutledge2002}
{Rutledge}, R.~E., {Bildsten}, L., {Brown}, E.~F., {Pavlov}, G.~G. \& {Zavlin}, V.~E. 2002, \apj, 577, 346

\bibitem[{{Str{\" u}der} {et~al.}(2001){Str{\" u}der}, {Briel}, {Dennerl},
  {Hartmann}, {Kendziorra}, {Meidinger}, {Pfeffermann}, {Reppin}, {Aschenbach},
  {Bornemann}, {Br{\" a}uninger}, {Burkert}, {Elender}, {Freyberg}, {Haberl},
  {Hartner}, {Heuschmann}, {Hippmann}, {Kastelic}, {Kemmer}, {Kettenring},
  {Kink}, {Krause}, {M{\" u}ller}, {Oppitz}, {Pietsch}, {Popp}, {Predehl},
  {Read}, {Stephan}, {St{\" o}tter}, {Tr{\" u}mper}, {Holl}, {Kemmer},
  {Soltau}, {St{\" o}tter}, {Weber}, {Weichert}, {von Zanthier},
  {Carathanassis}, {Lutz}, {Richter}, {Solc}, {B{\" o}ttcher}, {Kuster},
  {Staubert}, {Abbey}, {Holland}, {Turner}, {Balasini}, {Bignami}, {La
  Palombara}, {Villa}, {Buttler}, {Gianini}, {Lain{\' e}}, {Lumb}, \&
  {Dhez}}]{Strueder}
{Str{\" u}der}, L., {Briel}, U., {Dennerl}, K., {et~al.} 2001, \aap, 365, L18


\bibitem[{{Turner} {et~al.}(2001){Turner}, {Abbey}, {Arnaud}, {Balasini},
  {Barbera}, {Belsole}, {Bennie}, {Bernard}, {Bignami}, {Boer}, {Briel},
  {Butler}, {Cara}, {Chabaud}, {Cole}, {Collura}, {Conte}, {Cros}, {Denby},
  {Dhez}, {Di Coco}, {Dowson}, {Ferrando}, {Ghizzardi}, {Gianotti}, {Goodall},
  {Gretton}, {Griffiths}, {Hainaut}, {Hochedez}, {Holland}, {Jourdain},
  {Kendziorra}, {Lagostina}, {Laine}, {La Palombara}, {Lortholary}, {Lumb},
  {Marty}, {Molendi}, {Pigot}, {Poindron}, {Pounds}, {Reeves}, {Reppin},
  {Rothenflug}, {Salvetat}, {Sauvageot}, {Schmitt}, {Sembay}, {Short},
  {Spragg}, {Stephen}, {Str{\" u}der}, {Tiengo}, {Trifoglio}, {Tr{\" u}mper},
  {Vercellone}, {Vigroux}, {Villa}, {Ward}, {Whitehead}, \& {Zonca}}]{Turner}
{Turner}, M.~J.~L., {Abbey}, A., {Arnaud}, M., {et~al.} 2001, \aap, 365, L27


\bibitem[{Ueda}  {et~al.}(2005)]{Ueda2005}
{Ueda}, Y., {Ishisaki}, Y., {Takahashi}, T., {Makishima}, K. \& {Ohashi}, T. 2005, \apjs, 161, 185

\bibitem[{Vaughan}  {et~al.}(2003)]{Vaughan2003}
{Vaughan}, S. and {Edelson}, R. and {Warwick}, R.~S. and {Uttley}, P. 2003, \mnras, 345, 1271

\bibitem[{Willman}  {et~al.}(2002)]{Willman2002}
{Willman}, B., {Dalcanton}, J., {Ivezi{\'c}}, {\v Z}., {Jackson}, T., {Lupton}, R., {Brinkmann}, J. , {Hennessy}, G. \& 
{Hindsley}, R. 2002, \aj, 123, 848

\end{thebibliography}
\end{document}